\documentstyle [12pt]{article}
\begin{document}

\hfill {WM-00-108}

\hfill {\today}

\vskip 1in   \baselineskip 24pt
{
\Large 
   
   \bigskip
   \centerline{SCALAR MEDIATED FCNC AT THE FIRST MUON COLLIDER}}
 \vskip .4in
\def\bar{\overline}
 
\centerline{Marc Sher } 
\bigskip
\centerline {\it Nuclear and Particle Theory Group}
\centerline {\it Physics
Department}
\centerline {\it College of William and Mary, Williamsburg, VA
23187, USA}

\vskip .5in
 
{\narrower\narrower In the most general two-Higgs doublet model (generally referred
to as Model III), tree level scalar-mediated flavor changing neutral couplings exist. 
It has been noted that the most natural value for such a coupling is of the order of
the geometric mean of the Yukawa couplings of the two fermions.   Bounds on those
couplings that involve the second and third generations, from $\tau,B,D$ and $\mu$
physics, are very weak and are not sensitive to this ``natural" value.   In this
paper, it is pointed out that the process $\mu^+\mu^-\rightarrow \mu\tau$, at a muon
collider tuned to the scalar resonance, will easily reach this sensitivity if the
scalar mass is below $140$ GeV.  Hundreds of events are expected for an integrated
luminosity of an inverse femtobarn, and there appears to be no background.  Failure
to observe this process, if the scalar is below $140$ GeV, would effectively rule out
Model III.

\newpage

The most important unanswered question in the Standard Model(SM) is the nature of
electroweak symmetry breaking.   Should the Higgs bosons and/or additional scalars be
discovered in the next few years at either the Tevatron or the LHC, it will be
crucial to explore the properties of these particles in detail.  A primary motivation
for the muon collider\cite{fmc} is that a large number of scalars can be
produced through s-channel resonance, and thus a muon collider could be a ``Higgs"
factory.

The capabilities of a muon collider for exploring Higgs physics in the Standard Model
and in the Minimal Supersymmetric Standard Model(MSSM) have been explored
extensively\cite{fmc}.    The simplest extension of the Standard Model is the
two-Higgs doublet model(2HDM).    In this model, tree level flavor changing neutral
currents (FCNC) will naturally occur.  This is because there are two Yukawa coupling
matrices, and thus diagonalization of the quark mass matrix will not automatically
diagonalize each of the Yukawa matrices.   These FCNC are phenomenologically
dangerous, leading, for example,  to large contributions to $K^o-\overline{K}^o$
mixing.

It is important to note that one can always choose a basis in which one scalar field
acquires a vacuum expectation value and the other does not.  In this case, the latter
does not participate at all in electroweak symmetry breaking and is the only scalar
with FCNC couplings, and there is no {\it a priori} reason that its mass should be
light.  If its mass is greater than a few TeV\cite{ng}, there is no FCNC problem.
This is the most straightforward solution, and the effective theory below a TeV is
then just the standard model.

However, when people refer to the 2HDM, they generally refer to the
situation in which this extra scalar is light, with a mass on the order of the
electroweak scale.\footnote{This would involve fine-tuning, of course, but no more
fine-tuning than in the MSSM, where the SUSY breaking scale and the electroweak scale
are logically independent.}   In this case, something must suppress the
flavor-changing couplings.

One method of suppression\cite{gw}, which eliminates tree-level FCNC completely, is
to assume that a discrete symmetry either couples all of the fermions to only one of
the scalar doublets (Model I) or else couples one doublet to the Q=2/3 quarks and the
other to the Q=-1/3 quarks (Model II).  Such a discrete symmetry is completely {\it
ad hoc}.  Note that Model II type couplings automatically occur in the MSSM, but if an
additional pair of Higgs doublets is added to the MSSM, the same problem recurs.

Another method of suppression is to assume that the flavor-changing neutral
couplings are small.  It was pointed out in Ref. \cite{chengsher} that in a variety
of mass matrix models, the FCNC couplings are approximately given by the geometric
mean of the Yukawa couplings of the two fermions (this is now referred to as Model
III).    With this ansatz, the FCNC couplings involving the first generation fields
are very small, and the bounds are not as severe.   The largest flavor-changing
couplings will then involve those between the second and third generations.   The
FCNC couplings can only be bounded phenomenologically, and there are several detailed
analyses of these bounds\cite{sheryuan,ars5}, coming from
$K$,
$D$,
$B$ and
$\tau$ physics.

What will a muon collider be able to tell us about Model III?   Atwood, Reina and
Soni\cite{ars6} showed that the process $\mu^+\mu^-\rightarrow H\rightarrow
t\bar{c}+\bar{t}c$ could occur at an easily observable rate, since the geometric mean
of the top and charm Yukawa couplings is greater than the bottom quark Yukawa
coupling.   Depending on parameters, one could detect large numbers of events, with an
integrated luminosity of an inverse femtobarn, if the scalar mass exceed 180 GeV.

In this Letter, another signature of Model III, which is both cleaner than the
hadronic signature and which does not require the scalar to be heavier than the top
quark, is discussed.  This signature is
$\mu^+\mu^-\rightarrow H\rightarrow \mu\tau$.   For a 120 GeV scalar, for
example, one would see a $60$ GeV muon back-to-back with a $60$ GeV tau.  Since any
other process that produces a high-energy muon will result in a muon with less than
half the center-of-mass energy, there will be no background.  We will see that there
will be large number of events, and failure to detect this signal would virtually
rule out Model III.

Let us first consider the Yukawa interaction of two scalar doublets to fermions
\begin{equation}
{\cal L}=\eta_{ij}^U\bar{Q}_i\tilde{\phi}_1U_j+\eta_{ij}^D\bar{Q}_i\phi_1D_j
+\eta_{ij}^L\bar{L}\phi_1E_j+\xi_{ij}^U\bar{Q}\tilde{\phi}_2U_j+\xi_{ij}^D\bar{Q}
\phi_2D_j+\xi_{ij}^L\bar{L}\phi_2E_j
\end{equation}
In this Letter, we will consider the neutral  fields only, and will ignore
possible CP violation, focusing on the CP-even scalars.   

Without loss of generality, one chooses a basis such that only one scalar,
$\phi_1$, acquires a VEV,
$v=246$ GeV, and the other, $\phi_2$, does not (thus,
$\phi_2$ does not participate in symmetry breaking and does not really deserve
the label ``Higgs boson").   The Yukawa interactions of $\phi_1$ are 
proportional to the mass matrix and are therefore flavor-diagonal, whereas the Yukawa
interactions of $\phi_2$ are arbitrary and give FCNC.   In order to
prevent tree level FCNC, one must impose a discrete symmetry in Eq. 1, such as
$\phi_2\leftrightarrow -\phi_2$.  In Model III, no such symmetry is imposed.

For experimental purposes, one should use the physical mass
basis for the Higgs bosons, $h$ and $H$. We define the $h$ field to be the lighter of
the two.  In the conventional notation, this basis is rotated by an angle $\alpha$
from the basis of Eq. 1.  Ignoring the imaginary part of the Yukawa couplings, the
couplings of the
$h$ field to $\bar{f}_if_j$ is
\begin{equation}
C_{hf_if_j}= -{g\over 2}{m_i\over M_W}\delta_{ij}\sin\alpha +
{\xi_{ij}\over\sqrt{2}}\cos\alpha
\end{equation}

It was suggested by Cheng and Sher\cite{chengsher} that the most natural value for the
$\xi_{ij}$, which occurs in many mass-matrix models, is given by the geometric mean
of the Yukawa couplings of the two fermions, i.e. writing
\begin{equation}
{\xi_{ij}\over\sqrt{2}}=\lambda_{ij}{g\over 2}{\sqrt{m_im_j}\over m_W}
\end{equation}
one expects to have $\lambda_{ij}\simeq 1$.  

There have been a number of papers[4-15] examining Model III and
constraining the allowed values of $\lambda_{ij}$ (where  $i\ne j$).  The most
extensive analyses are those of Refs. \cite{sheryuan},\cite{ars5}.  In the former,
limits from rare
$\tau$ and rare
$B$ decays were considered, and in the latter, effects from $F^o-\bar{F}^o$ (where
$F=K,D,B,B_s$), $e^+e^- (\mu^+\mu^-) \rightarrow t\bar{c}+c\bar{t}$, $Z\rightarrow
b\bar{b}$, $t\rightarrow c\gamma$ and the $\rho$ parameter were considered.  The
result of these analyses show that $\lambda_{ds}^D<<1$, $\lambda_{db}^D<1$ and
$\lambda_{uc}^U<1$.  In other words, the FC couplings involving the first generation
seem to be very small.   

There are several 2HDM models\cite{dk,fefo} in which the $\lambda_{ij}$ involving the
first generation are much smaller than one, and one might argue that the extremely
small Yukawa couplings of the first generation might be subject to perturbations from
physics at a very high scale.  A true test of Model III would be to examine FC
couplings involving the second and third generations (if {\it those} are small, then
Model III would be excluded).  This might be particularly interesting in view of the
fact that the atmospheric neutrino problem indicates\cite{superk} very large mixing
between the second and third generations in the neutrino sector.   However, looking
at rare
$\tau$ and rare $B$ decays\cite{sheryuan} can not reach the $\lambda_{\mu\tau}\simeq
1$ or
$\lambda_{bs}\simeq 1$ sensitivity without an improvement of many orders of magnitude
over current limits, and $B_s-\bar{B}_s$ mixing is already maximal.  One can consider
$t\rightarrow c\gamma$, but the branching fractions are very small.  The decay
$t\rightarrow c h$ is promising\cite{hou}, although the branching fraction is less
than a percent (for
$\lambda_{ct}=1$).  Recently,
a very extensive analysis of detection of $t\rightarrow c h$ at the LHC has been
carried out\cite{asb}.   There it was shown that one would be able to detect the
decay for $\lambda_{ct}=1$ fairly easily, and that failure to detect the signal
would give $\lambda_{ct}<0.12$.  There has also been a detailed analysis\cite{bsesw}
of top-charm production at the NLC; although the signature is fairly clear due to the
distinct kinematics, the rate remains small.  

At a muon collider, one will be produce scalar bosons as an s-channel resonance, and
tens of thousands of scalars will be produced directly.  Atwood, Reina and
Soni\cite{ars6} showed how the process $h\rightarrow t\bar{c}+\bar{t}c$ would have a
very distinctive signature and a high event rate.  However, it does require the scalar
to have a mass above $180$ GeV.

Since large mixing has only been observed in the neutral lepton sector between the
second and third generations, one naturally would want to look at the charged lepton
sector, i.e. at $\lambda_{\mu\tau}$.   The muon collider provides a ``smoking gun"
signature for this coupling---the decay $h\rightarrow \mu\tau$.  This will have a
high event rate, and zero background.\footnote{This decay was considered in the
context of the Tevatron (where the scalar is produced off-resonance) in Ref.
\cite{dct}.  They showed that if the current bound on $\lambda_{\mu\tau}$ is
saturated, a signal of
$h\rightarrow\mu\tau$ could be seen at the Tevatron.}

The cross section for production of a state $X$ at the muon collider, convoluted with
the collider energy distribution,  is given by\cite{fmc}
\begin{equation}
\sigma_h \sim {4\pi\over m^2_h}{B(h\rightarrow\mu^+\mu^-)B(h\rightarrow X)\over
\big[ 1+{8\over \pi}\left( {\sigma_{\sqrt{s}}\over \Gamma_h}\right)^2\big]^{1/2}}
\end{equation}
where the $B$ are the branching ratios, $\Gamma_h$ is the total width and 
the Gaussian spread in the beam energy $\sqrt{s}$ is given by $\sigma_{\sqrt{s}} =
{R\over \sqrt{2}}\sqrt{s}$, with $R$ the energy resolution of each beam.  The energy
resolution is expected to be in the range $0.005\%-0.05\%$.  We will look at
relatively light scalars; for $m_h<140$ GeV, the Higgs boson is narrow and the cross
section is proportional to $\Gamma_h/R$.

The result depends on the couplings of the scalar to $\mu$ pairs as well as
$\mu-\tau$.  For the former, we will not include a possible $\xi_{\mu\mu}$
contribution--this is unknown and is expected to be of the same order as the standard
model contribution.  Note that any enhancement in this coupling will be determined as
soon as the resonance is found at the muon collider.    For the $\mu-\tau$
coupling, we will assume that $\lambda_{\mu\tau}\cos\alpha=1$ (the cross section will,
of course, scale as $\lambda_{\mu\tau}^2\cos^2\alpha$).   Since the total width of the
scalar boson is (for  masses below about $140$ GeV) dominated by $b$-quark
decays, we have
\begin{equation}
\Gamma_h={3g^2m_b^2m_H\over 32\pi m^2_W}
\end{equation} 
\begin{equation}
B(h\rightarrow \mu^+\mu^-) = {1\over 3}{m^2_\mu\over m^2_b},
\end{equation}
and
\begin{equation}
B(h\rightarrow \mu^+\tau^-) = {1\over 3}{m_\mu m_\tau\over m^2_b}
\end{equation}

For an integrated luminosity of 1 femtobarn, and a resolution of $R=0.005\%$ (both of
which are achievable), the total number of events is given by
$(930,740,540,330,100,1)$ for
$m_h=(100,110,120,130,140,150)$ GeV.   Note that for larger masses, the rate
drops considerably since new decay channels open up and the branching ratios drop
substantially.   The event rate also scales as $1/R$, should the given
resolution not be achievable.

We see that for  masses below $140$ GeV, we expect hundreds of events in an
inverse femtobarn.  The signal would be dramatic.  One
would see a muon and a tau each with an energy of half the center-of-mass energy.  
The impact parameter for tau decays at this energy is approximately 100
microns\cite{bhz}, which should be resolvable.  But even if the tau vertex can not be
seen, the mere presence of a muon with half the beam energy (and no muon of similar
energy on the other side) would indicate new physics, and the decays of the tau will
provide a smoking gun for this model.  There appears to be no substantial background;
any other particles produced which decay into muons will have the muon energy
degraded.

Suppose the scalar has a mass of $120$ GeV.  Then failure to see any signal
would put an upper bound on $\lambda_{\mu\tau}\cos\alpha$ of substantially less than
$0.1$ (and $\cos\alpha$ would likely be determined when the scalar is initially
discovered).  Since this is expected to be the largest coupling, such a bound
(coupled with failure to observe $t\rightarrow c h$ at the LHC\cite{asb}) would
effectively rule out Model III.  

What about other flavor-changing processes?  One can look for $\mu^+\mu^-\rightarrow
e\tau$ just as easily.  However, here the number of events expected (for
$\lambda_{e\tau}\cos\alpha=1$) is down by a factor of $m_\mu/m_e\sim 200$. Thus, one
would expect only a few events.  Nonetheless, the background is negligible, and this
would provide the best bound on $\lambda_{e\tau}$ (and would provide a good check if
the $\mu\tau$ signature is seen).   One can also look for $\mu^+\mu^-\rightarrow
\bar{b}s+b\bar{s}$, which would constitute roughly $3\%$ of all  decays, but the
background from $b\bar{b}$ would probably rule out substantive limits.

If the scalar mass is below $140$ GeV, then Model III predicts hundreds of $\mu\tau$
events will be observed at the first muon collider.  Although one can never
experimentally rule out tree level flavor changing neutral couplings at some level,
the primary motivation for Model III requires that the largest flavor changing
couplings give $\lambda\sim 1$.  The muon collider will thus provide a
definitive test of this Model.

I thank Vernon Barger and Lorenzo Diaz-Cruz for useful discussions, and Chris
Carone for reading the manuscript.  This work was supported the National Science
Foundation grant NSF-PHY-9900657.

\newpage

 \def\prd#1#2#3{{\rm Phys. ~Rev. ~}{\bf D#1} (19#2) #3}
\def\plb#1#2#3{{\rm Phys. ~Lett. ~}{\bf B#1} (19#2) #3 }
\def\npb#1#2#3{{\rm Nucl. ~Phys. ~}{\bf B#1} (19#2) #3 }
\def\prl#1#2#3{{\rm Phys. ~Rev. ~Lett. ~}{\bf #1} (19#2) #3 }

\bibliographystyle{unsrt}

\begin{thebibliography}{99}
\bibitem{fmc}
{\it Workshop on Physics at the First Muon Collider and at the Front End of the Muon
Collider}, eds. S. Geer and R. Raja (AIP Publishing, Batavia IL  1997); V. Barger,
M.S. Berger, J.F. Gunion and T. Han, Physics Reports {\bf 286} (1997) 1.
\bibitem{ng}
H. Georgi and D. V. Nanopoulos, \plb{82}{79}{95}.
\bibitem{gw}
S. Glashow and S. Weinberg, \prd{15}{77}{1958}.
\bibitem{chengsher}
T. P. Cheng and M. Sher, \prd{35}{87}{3484}.
\bibitem{sheryuan}
M. Sher and Y. Yuan, \prd{44}{91}{1461}.
\bibitem{ars5}
D. Atwood, L. Reina and A. Soni, \prd{55}{97}{3156}.
\bibitem{ars6}
D. Atwood, L. Reina and A. Soni, \prl{75}{95}{3800}.
\bibitem{ahr}
A. Antaramian, L. J. Hall and A. Rasin, \prl{69}{92}{1871}.
\bibitem{hw}
L. J. Hall and S. Weinberg, \prd{48}{93}{R979}.
\bibitem{savage}
M. J. Savage, \plb{266}{91}{135}.
\bibitem{lukesavage}
M. Luke and M. J. Savage, \plb{307}{93}{387}.
\bibitem{hou}
W.-S. Hou, \plb{296}{92}{179}.
\bibitem{ars12}
D. Atwood, L. Reina and A. Soni, \prd{53}{96}{1199}.
\bibitem{ars13}
D. Atwood, L. Reina and A. Soni, \prd{54}{96}{3295}.
\bibitem{dct}
J. L. Diaz-Cruz and J. J. Toscano, hep-ph/9910233.
\bibitem{asb}
J.A. Aguilar-Saavedra and G.C. Branco, hep-ph/0004190.
\bibitem{dk}
A. Das and C. Kao, \plb{372}{96}{106}.
\bibitem{fefo}
A. Aranda and M. Sher, hep-ph/0005113.
\bibitem{superk}
Y. Fukuda, et al., \prl{81}{98}{1562}.
\bibitem{bsesw}
S. Bar-Shalom, G. Eilam, A. Soni, J. Wudka, \prl{79}{97}{1217}.
\bibitem{bhz}
V. Barger, T. Han and C.-G. Zhao, hep-ph/0002042.
\end{thebibliography}

\end{document}